**Title:** Mapping the Intersection of Research and Policy in Centers for Medicare National Coverage Decision Memos

**Authors:** Sean A. Klein

**Author Affiliation:** Division of Program Coordination, Planning, and Strategic Initiatives; Office of the Director; National Institutes of Health; Bethesda, MD, USA

**Author E-Mail:** sean.klein@nih.gov

**Author ORCID:** 0000-0003-1960-3103

**Abstract**

Evidence is a crucial component of federal policy, but the interactions between the various stakeholders involved in funding, producing, and using the results of scientific research, an important class of evidence, for federal policy are poorly understood. The national coverage determination process used by the Centers for Medicare and Medicaid Services (CMS) to make significant policies on healthcare coverage is an ideal candidate for studying the interactions between stakeholders producing and utilizing scientific research for policy. Memos produced during the national coverage determination process contain information that identifies the organizations funding and producing research articles cited by CMS policy staff. I use these data to map scientific articles and their funding sources to discrete federal policies with substantial economic and health impacts. My analysis highlights that information derived from policy documents can facilitate transparency among the stakeholders involved in funding, producing, and using evidence for federal policy.

**Classification Code (JEL):** O32, O38, I18

**Introduction**

Evidence is a critical component of effective policy. Public Law 115 - 435 - Foundations for Evidence-Based Policymaking Act of 2018 Scientific research, an activity that generates evidence, is particularly important in health care, where policies must adapt to the accumulation of knowledge in various scientific fields (Mokyr, 2002). For example, governments made policy changes in response to the unpredictable nature of the COVID-19 pandemic including fluctuating case rates, new virus variants, and changes in the availability of therapeutics and personal protective equipment among many other factors.(Yin et al., 2021) Despite the importance of scientific research to federal healthcare policy, the ecosystem of stakeholders funding, conducting, and utilizing policy-relevant scientific research is not well understood. It remains challenging to answer questions like "which entities fund or conduct policy-relevant research" or "how strong are the connections among stakeholders".(Yin et al., 2022)

Medicare national coverage determination decision memos ("memos") represent a unique opportunity to study how scientific research is used in policy since they contain consistently formatted references that are amenable to automated data extraction across all memos. The national coverage determination process dictates which technologies and services are nationally covered by Medicare and has important implications for the 34 million Medicare beneficiaries and the federal budget as Medicare spending was $900 billion in 2020 (CMS, 2024a). The memos are written by the Centers for Medicare & Medicaid Services (CMS) policy staff with subject matter expertise in clinical and scientific research and are made public to promote

transparency. Memos provide a consistently formatted list of references that link scientific research to coverage decisions.

Though CMS policy staff produce the memos, they generally cite published research that is independently funded and conducted.[1] CMS policy staff extensively review the content (i.e., methods, analyses, conclusions) of the articles they cite when writing memos and other studies have examined the use of evidence in memos but no group has attempted to analyze the aggregate contributions of the stakeholders funding and producing the research cited across all memos, which I will call the "memo research ecosystem" (Chambers et al., 2015). I propose that the contributions of stakeholders can be measured using the reference sections in the memos. I also propose that mapping the memo research ecosystem would make it easier for funders and scientists to identify when their research is used to inform policy linked to important healthcare and federal spending outcomes, while CMS could use the map to facilitate engagement with funders and scientists on topics of interest to the Medicare program.

In this article, I describe how to map the memo research ecosystem using linkages between memo references, articles, and funding data. I create a database by linking memos retrieved from the CMS website to articles indexed in PubMed, a biomedical literature database (NLM, 2024). I then use funding data included in the articles to identify the funders and research organizations that make up the research ecosystem.

---

[1] Notable exceptions where CMS does cover some expenses related to clinical trial participation are the Clinical Trial Policies (CMS, 2024c), Coverage with Evidence Development (CMS, 2014), and Investigation Device Exemption Studies (CMS, 2024b).

Finally, I demonstrate analyses that the research ecosystem makes possible, including identifying relevant stakeholders and estimating the magnitude of their contributions. I find that the aggregated memo research ecosystem consists of a small number of dominant funders supporting many research organizations (i.e., institutions conducting research, such as universities) and that the structure of the ecosystem varies depending on the subject matter of the memo.

**Study Data And Methods**

*Policy documents and processing*

All memos were manually retrieved during the month of March, 2020 from the CMS website as PDFs and their reference sections were extracted using custom scripts (Anaconda, 2020; Shinyama Y, 2019). A custom python script matched the text from each reference section to articles indexed in PubMed using lexical, a citation resolution software (Lexical, 2021). A custom python script utilizing the National Institutes of Health (NIH) E-Utilities functions was used for any text that could not be assigned by lexical but only identified an additional five (>1% of the total) articles (NCBI, 2018). The resulting dataset contained PubMed reference numbers (PMIDs) associated with the citing memo's title and its publication date.

*Article-Award Linkage*

The PMIDs were used to search PubMed (using a custom Python script and NIH E-Utilities) for any awards (i.e., grant or contract) listed in the article or to search NIH RePORTER (using an R package for the RePORTER application programming interface) for federal awards that cited the PMID (Belter, 2021; NIH, 2024). PubMed data was used to identify non-federal awards and only contained award identification numbers and the funding organization ("funders"). Federal award data from RePORTER provided award identification numbers, funders, and awardee organization ("research organization"). Funder names from PubMed and RePORTER were manually standardized to the RePORTER format. DUNS codes provided by RePORTER were used to uniquely identify research organizations.

*Analysis of funding data*

The availability of detailed funding information for NIH Institutes and Centers ("ICs") allowed me to estimate whether memos relied more heavily on research funded by specific ICs. I created a memo funding set composed of awards associated with memo articles and funded by ICs and an all-NIH funding set composed of all the awards issued by NIH ICs from 1985-2020. Assigning the cash value of an award to a publication was challenging (e.g., when multiple publications cited the same award) so I opted to use the number of awards rather than their cash value as a measure of an IC's contribution to funding for memo research and NIH research. The all-NIH funding and memo funding sets were respectively aggregated by the funding IC and award year (see Supplement for details). The difference in the percentage of awards funded by an IC for a specific year was calculated as:

$$(1) \quad \text{Difference}_{i,t} = \text{Memo}_{i,t} - \text{NIH}_{i,t}$$

Where $Memo_{i,t}$ represents the percentage of awards cited by all memo articles (not just those in the memo funding set) and funded by IC $i$ in year $t$, $NIH_{I,t}$ represents the percentage of all NIH awards funded by IC $i$ in year $t$, and $Difference_{i,t}$ represents the difference between the two for an IC-year combination. The distribution of differences for each IC was then tested for significant deviation from zero using a Wilcox rank sum test. ICs with fewer than five observations (i.e., funded at least one award in fewer than five years) were excluded from the analysis.

**Study Results**

Of the 225 decision memos available from the CMS database in May 2021, I capture 146 (65%) memos spanning 21 years (1999-2020) with reference section text that links to at least one article in PubMed. In most memos, a large proportion of reference text is linked to an article (Supplementary Figure 1, median = 79% of text, interquartile range = 25%) with 100% linkage being unlikely due to memos citation of material not indexed in PubMed (e.g., Food and Drug Administration [FDA] guidances or websites). 6,457 articles spanning 71 years (1948-2019) are associated with memo reference text. Of these, funding sources spanning 42 years (1979-2021) can be identified for 1,110 (17%) articles. 2,742 unique awards can be identified with 2,031 (74%) reporting research organization information. Fifty-eight funders supported these awards with NIH ICs funding the largest proportion (Table 1). Three ICs account for about half the total article-award pairs in the data (NHLBI, NCI, NIA; see Table 1 for abbreviation definitions). By comparison, awards to the 263 unique research organizations are much more evenly distributed with the most well-funded recipient receiving only 3.3% of the total number of awards and less than a quarter of all awards going to a top-ten recipient research organization (Table 2).

The median difference columns show the difference in an IC's (Table 1) or a research organization's (Table 2) share of NIH funded awards in memos versus all NIH awards. In Table 1 for example, NHLBI has a median difference of 14%, indicating that NHLBI funds about 14% more of the awards cited in memos than it funds for all NIH awards in any given year. Memos appear to cite more research funded by NCI, NHLBI, and NIAID and less research funded by NIAID than would be expected based on their shares of all NIH awards (see Supplemental Table 1 for additional ICs). For the award recipients listed in Table 2, median difference values were more modest (max median difference = 3) though still significant for many research organizations (Supplemental File 1).

Table 1

**Ten largest funders of CMS-cited research**

| Funder | # Awards | % Awards | Median Difference (%) | p-value |
|---|---|---|---|---|
| NCATS | 200 | 7 | 6 (4, 10) | 3E-09 |
| NCI | 566 | 21 | 11 (7, 14) | 7E-08 |
| NHLBI | 552 | 20 | 14 (10, 16) | 2E-09 |
| NIA | 335 | 12 | 7 (6, 14) | 4E-09 |
| NIAID | 79 | 3 | -6 (-7, -4) | 6E-03 |
| NIAMS | 58 | 2 | 1 (0, 2) | 3E-02 |
| NIDDK | 234 | 8 | 0 (-1, 8) | 2E-01 |
| NIMH | 89 | 3 | -2 (-2, 0) | 2E-02 |
| NINDS | 108 | 4 | -2 (-3, 0) | 1E-01 |
| PHS | 86 | 3 | - | - |

**Table 2**

**Ten largest recipients of awards cited in memo articles**

| Recipient | # Awards | % Awards | Median Difference | | p-value |
|---|---|---|---|---|---|
| Duke University | 46 | 1.7 | 1 | (1, 3) | 1E-04 |
| Johns Hopkins University | 92 | 3.4 | 2 | (1, 3) | 5E-06 |
| Massachusetts General Hospital | 49 | 1.8 | 2 | (1, 4) | 2E-04 |
| Mayo Clinic Rochester | 68 | 2.5 | 2 | (2, 4) | 2E-06 |
| University of California Los Angeles | 64 | 2.3 | 1 | (1, 3) | 4E-05 |
| University of California San Francisco | 58 | 2.1 | 1 | (1, 3) | 3E-04 |
| University of Pennsylvania | 44 | 1.6 | 1 | (0, 3) | 4E-03 |
| University of Pittsburgh | 85 | 3.1 | 2 | (1, 4) | 1E-06 |
| University of Washington | 86 | 3.1 | 3 | (2, 3) | 3E-07 |
| Washington University | 41 | 1.5 | 1 | (0, 2) | 3E-04 |

**Notes.** The percent of awards is calculated as the award number over the total number of awards (N = 2,742). Median difference is calculated using equation 1 with bracketed numbers indicating the bounds of a 95% confidence interval. Positive values indicate the research organization is overrepresented in memo article funding while negative values indicate it is underrepresented. P-values are estimated from a Wilcoxon signed rank test.

Figure 3 visualizes the memo research ecosystem for two memos. The research ecosystem for the memo on the use of beta amyloid PET in detecting dementia and neurodegeneration has a single dominant funder (NIA, 78% of awards) and a small number of research organizations used to inform the memo (55% of awards are to the top ten research organizations). In contrast, the research ecosystem of the memo on erythropoietin stimulating agents has a more diverse set of funders (NCI 43%; NIDDK 25%; NHLBI 13% of awards) and research organizations (38% of awards are to the top ten research organizations). The topology of the research ecosystem varies from memo to memo ranging from highly concentrated on a subset of funders and/or research organizations to relatively equal participation by all funders and/or research organizations (Supplemental Table 2, Supplemental Methods). However, further interpretation of topology is challenging due to the lack of completeness in funding information for the memo articles (only 17% could be connected back to funding sources).

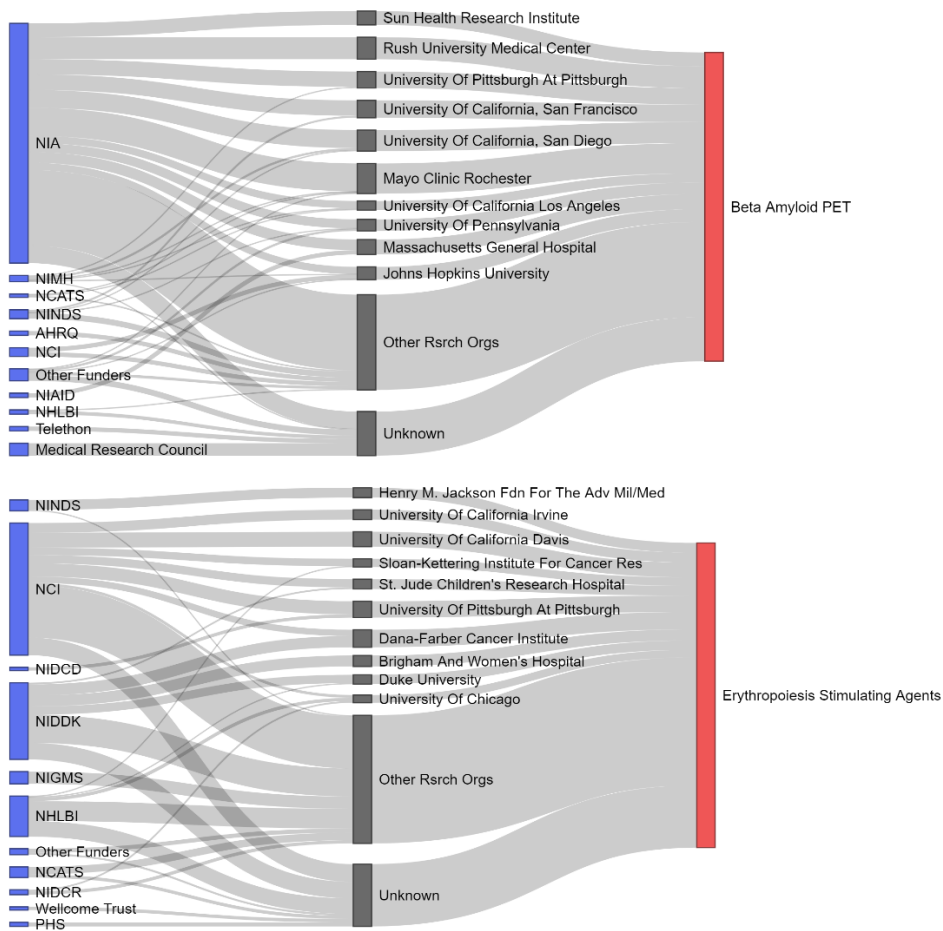

**Fig. 1** Model of research flows to two CMS policy memo. Funding institutions are blue blocks to the left, award recipients are gray blocks in the center, and the CMS memo is the red block to the right. Gray ribbons connecting stakeholders represent the proportion of memo articles that entity supported. Memo articles citing multiple awards were divided equally amongst the participating stakeholders. Only the top ten research organizations by proportion of memo articles are named with all other organizations aggregated as "Other". Awards cited in memo articles that couldn't be linked to a research organization are aggregated under the Unknown category.

**Discussion**

I find that the references in CMS decision memos ("memos") are a rich resource that can be used to identify and, to a limited extent, quantify the interactions among the stakeholders involved in funding, producing, and using research for policy. A critical component of this analysis is the use of automation and expansive article/funding databases, an approach leveraged in several other studies of policy documents and that here allowed for investigating all available memos rather than a subset (Drew et al., 2016; Haunschild & Bornmann, 2017). A notable benefit of capturing article citations in memos is that it simplifies evaluation of an important class of evidence used to inform policy. For example, articles cited in policy documents may be retracted long after the policy is published. A policy article database could facilitate the scientific integrity of federal policies by making it easier to identify policies that cite retracted articles (*Protecting the Integrity of Government Science*, 2022).[2]

Besides improving the scientific integrity of federal policy, indexing the articles cited in policy documents reduces barriers to rewarding funders and scientists for producing evidence that is useful to making important policy decisions. Scientists face strong incentives to publish highly cited articles but policy documents (e.g., agency regulations or guidance) are usually not indexed in citation counts.[3] Further, simply including policy citations among article citations would make little impact because articles are cited far more frequently by other articles than policy documents. Indexing citations to policy decisions with significant impacts on the public, like the memos used in this analysis, is one option for rewarding scientists producing policy-relevant research. Currently, these researchers may be undervalued: An expansive analysis of scientific articles used in the public domain found that federal documents tend to cite research that is not among the 1% most highly cited articles.(Yin et al., 2022) Considering policy citations as an additional dimension of impact could therefore elevate the importance of research that might be less well-cited by scientists but valuable to policy makers.

---

[2] I thank Stephen G. Gonsalves, who proposed this idea and approach.

Quantifying the memo research ecosystem permits analysis of the alignment among stakeholders, though results must be cautiously interpreted due to the low coverage of funding data (17% of articles). Unsurprisingly, the most aligned funders of the research cited in memos (e.g., NHLBI, NCI, NIA) have scientific missions that address common conditions in the Medicare population (cardiovascular diseases; cancer; and aging-related diseases like neurodegeneration) (CMS, 2022). The memo research ecosystem is useful because it identifies specific technologies and services where funders and CMS staff have shared interest. For example, the research ecosystem for the Positron Emission Tomography (PET) amyloid beta memo identifies the largest funder (NIA, 132 awards) and stratifies its awards by the number of CMS-cited articles they fund. Awards with multiple CMS-cited articles (N = 40) may reflect stronger alignment of interests between NIA and CMS. Several of these awards are cooperative agreements (Alzheimer's Disease Cooperative Study and Alzheimer's Disease Patient Registry) where NIA has a greater ability to direct the kinds of research conducted under the award. These data could be useful in guiding discussions between NIA and CMS about research on imaging for neurodegenerative disease. The NIA example is not unique and a similar approach can be used with any memo where funding data exist to identify shared interests among CMS and those funders or research organizations.

---

[3] Some notable exceptions are third party databases that index a subset of policy documents such as Altmetric (Altmetric, 2024) and Overton (Overton, 2024).

The memo research ecosystem also identifies non-obvious connections among stakeholders for specific topic areas. One example is detecting or treating infectious disease, which is particularly relevant in light of evolving risks from emerging pathogens and antimicrobial resistance (Baker et al., 2022; CARB, 2020). The research ecosystem for the three infectious disease-focused memos shows that NIAID, PHS, and NIDA are the largest funders and may be the most relevant partners in this space. NIDA's prominence is surprising given their IC's mission (research on biological, behavioral, and social implications; consequences; prevention; and treatment of drug use) does not directly relate to the diseases mentioned in the memos (Hepatitis C, HIV, and ulcers caused by *H. pylori*). Closer examination of the funding data shows that NIDA supported thirteen separate awards focused on HIV, AIDS, and Hepatitis C, mostly aimed at understanding these diseases in the context of drug use. This example demonstrates how the memo research ecosystem can be used to identify specific components of a funder's broader portfolio with substantial relevance to CMS policy makers, even when that connection is not obvious from the funder's mission.

Several limitations should be considered when interpreting my results. First, the data are incomplete both in terms of memos (65.8% of memos captured) and funding data (17% of articles had funding data). The lack of funding data is surprising given the high proportion of article-award linkages observed for other studies.(Azoulay et al., 2019; Cleary et al., 2018) It is unclear whether this is an artifact of the analysis, a feature of the data (e.g., many articles are funded by industry), or some combination thereof. Regardless, data scarcity makes it difficult to identify stakeholders and measure their contributions to the memo research ecosystem. An additional issue with my approach is that it uses only the published literature indexed in PubMed, which is a subset (~79%, see Results) of the information relied on by CMS policy staff during the national coverage decision process. Part of the remaining 21% of reference information may be available in more comprehensive literature databases (e.g., Web of Science, Scopus, Embase) but complete coverage of memo citations requires data from outside these resources. Documents not indexed in literature databases, such as guidances or regulatory review documents, may play an important role that I do not capture in my analysis. Despite these issues, I believe my analysis offers a reasonable first approximation of a research ecosystem underlying an important set of policy documents.

**Conclusion**

My work demonstrates how the memo research ecosystem can be constructed from existing references and open access data, and highlights some of the analyses that could be useful to various stakeholders. Delineating the connections between the policy and research spheres can provide useful benefits: policy makers can be better informed about their partners in the research community while funders and scientists can better assess the impact of their research and be rewarded appropriately. Additionally, all parties benefit from the increased visibility of the evidence used in policy, supporting scientific integrity efforts (Biden, 2021). The method I present here offers a data-driven approach to fostering relationships among the diverse stakeholders involved in funding, producing, and using evidence during the policy development process.

**Supplement**

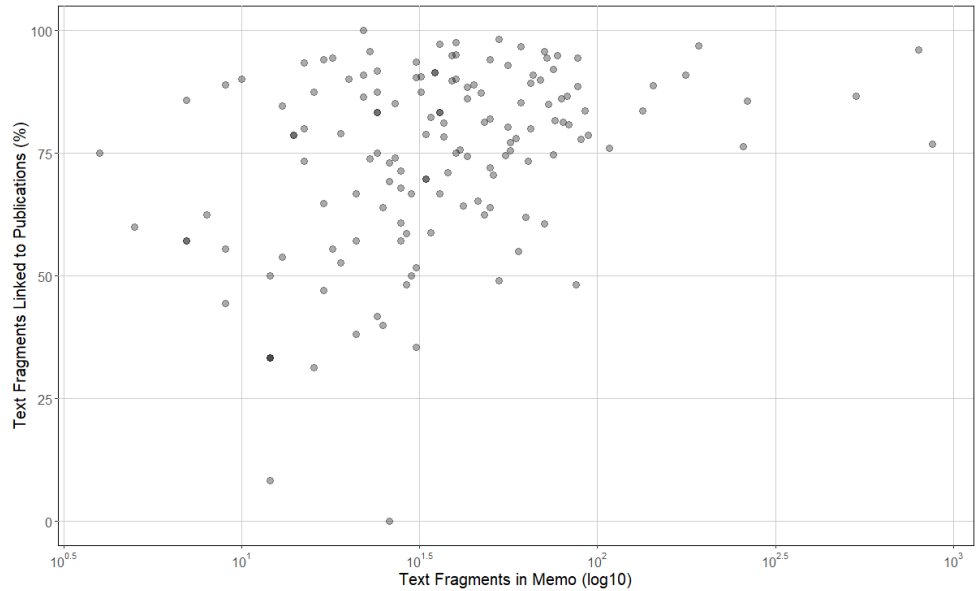

**Supplementary Figure 1. Relationship between amount of memo text and percentage of memo text linked to a publication.** Text fragments are free text extracted from memo reference sections and are checked to see if they match a publication in PubMed (see Methods). The x-axis is on $\log_{10}$ scale. Points represent memos (N = 146).

**Supplementary Table 1**

**Percentage and Difference of Shares of Funders Supporting Memo Articles**

| Funder | Awards (#) | Awards (%) | N | Median Difference | p-value |
|---|---|---|---|---|---|
| Agency for Healthcare Research & Quality | 38 | 1.39 | | | |
| Arthritis Research UK | 2 | 0.07 | | | |
| Agency for Toxic Substances & Disease Registry | 2 | 0.07 | | | |
| Austrian Science Fund FWF | 1 | 0.04 | | | |
| Bureau of Health Professions | 1 | 0.04 | | | |
| Brazilian National Council for Scientific and Technological Development | 2 | 0.07 | | | |
| British Heart Foundation | 10 | 0.36 | | | |
| Canadian Institutes of Health Research | 19 | 0.69 | | | |
| Cancer Research UK | 5 | 0.18 | | | |
| Chief Scientist Office | 12 | 0.44 | | | |
| CLC | 4 | 0.15 | | | |
| DS | 1 | 0.04 | | | |
| Fundação de Apoio à Pesquisa do RN | 1 | 0.04 | | | |
| U.S. Food & Drug Administration | 6 | 0.22 | | | |
| FIC | 2 | 0.07 | | | |
| The São Paulo Research Foundation | 1 | 0.04 | | | |
| U.S. Centers for Disease Control & Prevention | 2 | 0.07 | | | |
| Intramural NIH | 20 | 0.73 | | | |
| Medical Research Council | 51 | 1.86 | | | |
| Ministry of Health and Welfare | 1 | 0.04 | | | |
| Ministry of Science and Technology, Taiwan | 3 | 0.11 | | | |
| Multiple Sclerosis Society | 2 | 0.07 | | | |
| NCATS | 200 | 7.29 | 33 | 6 (4.5, 9.6) | 3E-09 |
| NCCAM | 7 | 0.26 | | | |
| NCCDPHP | 2 | 0.07 | | | |
| NCCIH | 2 | 0.07 | | | |

| | | | | | | |
|---|---|---|---|---|---|---|
| NCHHSTP | 2 | 0.07 | | | | |
| NCI | 566 | 20.64 | 35 | 11 | (6.8, 14.3) | 7E-08 |
| NCMHD | 3 | 0.11 | | | | |
| NEI | 28 | 1.02 | 12 | 0 | (-1.2, 2.2) | 1E+00 |
| NHGRI | 1 | 0.04 | | | | |
| NHLBI | 552 | 20.13 | 34 | 14 | (10.1, 16.4) | 2E-09 |
| NIA | 335 | 12.22 | 30 | 7 | (5.9, 13.7) | 4E-09 |
| NIAID | 79 | 2.88 | 16 | -6 | (-7.1, -4) | 6E-03 |
| NIAMS | 58 | 2.12 | 23 | 1 | (0.2, 2.1) | 3E-02 |
| NIBIB | 13 | 0.47 | 10 | 0 | (-0.5, 0.4) | 6E-01 |
| NICHD | 23 | 0.84 | 11 | -3 | (-3.2, -2.3) | 1E-03 |
| NIDA | 45 | 1.64 | 14 | -1 | (-1.5, 0.4) | 3E-01 |
| NIDCD | 4 | 0.15 | | | | |
| NIDCR | 7 | 0.26 | 6 | 0 | (-0.5, 1.1) | 6E-01 |
| NIDDK | 234 | 8.53 | 33 | 0 | (-0.8, 8) | 2E-01 |
| NIEHS | 13 | 0.47 | 9 | 0 | (-0.9, 0.8) | 9E-01 |
| NIGMS | 36 | 1.31 | 19 | -8 | (-8.2, -5.4) | 4E-06 |
| NIH | 2 | 0.07 | | | | |
| NIMH | 89 | 3.25 | 24 | -2 | (-2.3, -0.3) | 2E-02 |
| NIMHD | 5 | 0.18 | | | | |
| NINDS | 108 | 3.94 | 27 | -2 | (-2.6, 0.3) | 1E-01 |
| NINR | 8 | 0.29 | 8 | 0 | (0, 2.1) | 3E-02 |
| NIOSH | 1 | 0.04 | | | | |
| NLM | 4 | 0.15 | | | | |
| Office of Chief of Public Health Practice | 1 | 0.04 | | | | |
| PHS | 86 | 3.14 | | | | |
| Coord. for the Improvement of Higher Education Personnel Nat. Postdoc. Program | 1 | 0.04 | | | | |
| SingHealth Foundation Research Grant | 1 | 0.04 | | | | |

| | | |
|---|---|---|
| Telethon | 1 | 0.04 |
| UK Department of Health | 17 | 0.62 |
| U.S. Dept. of Veterans Affairs | 2 | 0.07 |
| Wellcome Trust | 20 | 0.73 |

Notes. The percent of awards is calculated as the award number over the total number of awards (N = 2,742). Median difference is calculated using equation 1 with bracketed numbers indicating the bounds of a 95% confidence interval. Positive values indicate the IC is overrepresented in the memo data while negative values indicate it is underrepresented. P-values are estimated from a Wilcoxon signed rank test. Acronyms in alphabetical order: National Center for HIV, Viral Hepatitis, STD, and Tuberculosis Prevention (NCHHSTP), National Institute for Occupational Safety and Health (NIOSH), Public Health Services (PHS), all other acronyms are NIH Institutes, Centers, or Offices that can be found at (NIH, 2022).

**Supplemental Table 2**

**Differences in article preferences between CMS policy staff and scientists by memo.**

| Memo Title | # Memo Articles w/ RCRs | Median $\Delta RCR$ | | # Memo Articles w/ Year | Median $\Delta Year$ | | $KLD_F$ | $KLD_{RO}$ |
|---|---|---|---|---|---|---|---|---|
| Decision Memo for Ambulatory Blood Pressure Monitoring (ABPM) (CAG-00067R2) | 25 | 3.3 | (0.9, 30) | 25 | -3 | (-4, -1) | 0.5 | 0.6 |
| Decision Memo for Arthroscopy for the Osteoarthritic Knee (CAG- 00167N) | 30 | -0.8 | (-1.4, 0.9) | 30 | 1 | (-4, 3) | 0.1 | 0.1 |
| Decision Memo for Autologous Blood-Derived Products for Chronic Non-Healing Wounds (CAG-00190N) | 41 | -0.7 | (-1.1, -0.1) | 41 | -2 | (-3, 1) | 0.1 | 0.1 |
| Decision Memo for Autologous Blood-Derived Products for Chronic Non-Healing Wounds (CAG-00190R3) | 64 | 0 | (-0.4, 0.8) | 64 | -1 | (-3, 1.5) | 0.0 | 0.0 |
| Decision Memo for Autologous Stem Cell Transplantation (AuSCT) for Multiple Myeloma (CAG-00011N) | 20 | 0.1 | (-0.3, 2.2) | 20 | -2 | (-3.5, -1.5) | 0.0 | 0.2 |
| Decision Memo for Bariatric Surgery for the Treatment of Morbid Obesity - Facility Certification Requirement (CAG-00250R3) | 28 | -0.4 | (-0.8, 0.2) | 28 | -2 | (-3, 0) | 0.5 | 0.7 |
| Decision Memo for Beta Amyloid Positron Emission Tomography in Dementia and Neurodegenerative Disease (CAG-00431N) | 85 | 4.8 | (3.6, 7.5) | 85 | -4 | (-4, -3) | 1.7 | 0.4 |
| Decision Memo for Blood Brain Barrier Disruption (BBBD) Chemotherapy (CAG-00333N) | 39 | 0 | (-0.7, 1.3) | 40 | 6 | (4, 8.5) | 0.3 | 0.9 |
| Decision Memo for Carotid Artery Stenting (CAG-00085R) | 23 | 9.6 | (5, 23.1) | 23 | -1 | (-3, -1) | 0.3 | 0.2 |
| Decision Memo for Chimeric Antigen Receptor (CAR) T-cell Therapy for Cancers (CAG-00451N) | 30 | 13 | (6.4, 23.2) | 30 | -1 | (-2, 2) | 0.9 | 0.7 |
| Decision Memo for Collagen Meniscus Implant (CAG-00414N) | 48 | 0.6 | (-0.1, 1.4) | 51 | -1 | (-3, 2) | 0.0 | 0.0 |

| Decision Memo | | | | | | | | |
|---|---|---|---|---|---|---|---|---|
| Decision Memo for Computed Tomographic Angiography (CAG- 00385N) | 44 | 1 | (-0.6, 3.7) | 45 | -4 | (-4, -4) | 0.1 | 0.0 |
| Decision Memo for Continuous Positive Airway Pressure (CPAP) Therapy for Obstructive Sleep Apnea (OSA) (CAG-00093R) | 24 | -1 | (-1.5, 0.2) | 24 | -1 | (-2, 1) | 0.5 | 0.4 |
| Decision Memo for Continuous Positive Airway Pressure (CPAP) Therapy for Obstructive Sleep Apnea (OSA) (CAG-00093R2) | 62 | 0.6 | (0.1, 1) | 62 | -1 | (-2.5, 1) | 0.1 | 0.1 |
| Decision Memo for Electrical Bioimpedance for Cardiac Output Monitoring (CAG-00001R) | 90 | -1.2 | (-1.4, -0.9) | 91 | 0 | (-1, 1) | 0.1 | 0.1 |
| Decision Memo for Electrodiagnostic Sensory Nerve Conduction Threshold (CAG-00106R) | 36 | -1.5 | (-1.8, -0.9) | 36 | 0 | (-4.5, 1) | 0.0 | 0.0 |
| Decision Memo for Electrostimulation for Wounds (CAG-00068N) | 34 | -0.3 | (-1, 0.4) | 40 | 2.5 | (1, 6) | 0.1 | 0.0 |
| Decision Memo for Erythropoiesis Stimulating Agents (ESAs) for non-renal disease indications (CAG-00383N) | 498 | -0.6 | (-0.8, -0.4) | 503 | 2 | (1, 3) | 0.9 | 0.6 |
| Decision Memo for External Counterpulsation (ECP) Therapy (CAG- 00002R2) | 44 | -2 | (-2.2, -1.7) | 46 | -3 | (-5, -2.5) | 0.0 | 0.0 |
| Decision Memo for Extracorporeal Photopheresis (CAG-00324R) | 34 | 0.4 | (-0.2, 1.2) | 35 | -1 | (-3, 2) | 0.2 | 0.1 |
| Decision Memo for Extracorporeal Photopheresis (ECP) (CAG- 00324R2) | 37 | 1 | (-0.1, 2.2) | 39 | 6 | (1, 7) | 0.2 | 0.1 |
| Decision Memo for Ferrlecit®: Intravenous Iron Therapy (CAG- 00046N) | 24 | -0.1 | (-1.5, 1.1) | 24 | -2 | (-4, -1.5) | 0.0 | 0.0 |
| Decision Memo for Gender Dysphoria and Gender Reassignment Surgery (CAG-00446N) | 301 | -0.1 | (-0.3, 0.3) | 335 | 0 | (-1, 1) | 0.3 | 0.2 |
| Decision Memo for Heartsbreath Test for Heart Transplant Rejection (CAG-00394N) | 36 | -0.3 | (-0.6, 0.4) | 36 | 5.5 | (3, 8) | 0.2 | 0.6 |
| Decision Memo for Implantable Cardioverter Defibrillators (CAG- 00157R4) | 43 | 3.8 | (1.5, 10.1) | 43 | 4 | (1, 7) | 0.7 | 0.3 |
| Decision Memo for Infrared Therapy Devices (CAG-00291N) | 158 | -0.3 | (-0.8, 0) | 159 | 0 | (-1, 1) | 0.5 | 0.2 |

| Decision Memo | | | | | | | | |
|---|---|---|---|---|---|---|---|---|
| Decision Memo for Intestinal and Multivisceral Transplantation (CAG-00036N) | 23 | 0.9 | (-0.6, 4) | 23 | -3 | (-5, -2) | 0.0 | 0.0 |
| Decision Memo for Leadless Pacemakers (CAG-00448N) | 31 | -0.1 | (-1, 2) | 31 | -4 | (-4, -4) | 0.0 | 0.0 |
| Decision Memo for Lumbar Artificial Disc Replacement (CAG- 00292N) | 36 | 1.5 | (-0.3, 3.9) | 37 | -4 | (-5, -4) | 0.0 | 0.0 |
| Decision Memo for Lumbar Artificial Disc Replacement (LADR) (CAG-00292R) | 28 | 0.3 | (-0.8, 1.2) | 28 | -5 | (-6, -3) | 0.1 | 0.0 |
| Decision Memo for Magnetic Resonance Angiography of the Abdomen and Pelvis (CAG-00142N) | 20 | -0.2 | (-0.9, 0.3) | 20 | -2 | (-3, -1.5) | 0.2 | 0.0 |
| Decision Memo for Magnetic Resonance Imaging (MRI) (CAG- 00399R) | 32 | 0.3 | (-1, 1.3) | 32 | -1 | (-2, 3) | 0.2 | 0.1 |
| Decision Memo for Magnetic Resonance Imaging (MRI) (CAG- 00399R2) | 27 | 1.4 | (-0.5, 4.7) | 27 | -2 | (-3, -1) | 0.4 | 0.1 |
| Decision Memo for Magnetic Resonance Imaging (MRI) (CAG- 00399R4) | 27 | -0.1 | (-1, 0.8) | 27 | -4 | (-5, -2) | 0.5 | 0.4 |
| Decision Memo for Microvolt T-wave Alternans (CAG-00293N) | 20 | 0.9 | (0.1, 2.4) | 20 | -3 | (-4, -2) | 0.3 | 0.0 |
| Decision Memo for Microvolt T-wave Alternans (CAG-00293R) | 23 | 0.1 | (-0.3, 0.7) | 24 | -3 | (-4.5, -1) | 0.2 | 0.1 |
| Decision Memo for Microvolt T-wave Alternans (CAG-00293R2) | 40 | -0.7 | (-1.2, 0.5) | 41 | -1 | (-2, 0) | 0.4 | 0.5 |
| Decision Memo for Neuromuscular Electrical Stimulation (NMES) for Spinal Cord Injury (CAG-00153R) | 25 | -0.4 | (-1, -0.2) | 25 | -2 | (-3, -1) | 0.0 | 0.0 |
| Decision Memo for Next Generation Sequencing (NGS) for Medicare Beneficiaries with Advanced Cancer (CAG-00450R) | 27 | -0.8 | (-1.1, 0.1) | 27 | -3 | (-3, -2) | 0.4 | 0.1 |
| Decision Memo for Ocular Photodynamic Therapy (OPT) with Verteporfin for Macular Degeneration (CAG-00066R4) | 34 | -0.6 | (-1, 0.4) | 34 | 0 | (-1, 1) | 0.0 | 0.1 |

| Decision Memo | n | | | n | | | | |
|---|---|---|---|---|---|---|---|---|
| Decision Memo for Percutaneous Image-guided Lumbar Decompression for Lumbar Spinal Stenosis (CAG-00433N) | 24 | -1.6 | (-1.9, -0.9) | 24 | -4 | (-6, -3) | 0.3 | 0.4 |
| Decision Memo for Percutaneous Image-guided Lumbar Decompression for Lumbar Spinal Stenosis (CAG-00433R) | 40 | -1.1 | (-1.6, -0.2) | 40 | -3 | (-4, -1.5) | 0.2 | 0.3 |
| Decision Memo for Percutaneous Left Atrial Appendage (LAA) Closure Therapy (CAG-00445N) | 101 | 2.6 | (1.8, 4.8) | 102 | -1 | (-2, 0) | 0.5 | 0.3 |
| Decision Memo for Percutaneous Transluminal Angioplasty (PTA) and Stenting of the Renal Arteries (CAG-00085R4) | 20 | 0 | (-1.7, 7.7) | 20 | -1 | (-3, 1) | 0.2 | 0.1 |
| Decision Memo for Percutaneous Transluminal Angioplasty (PTA) of the Carotid Artery Concurrent with Stenting (CAG-00085N) | 44 | 2.8 | (1.8, 4.5) | 45 | -3 | (-3, -1) | 0.4 | 0.1 |
| Decision Memo for Positron Emission Tomography (FDG) (CAG- 00065N) | 40 | 1 | (0.2, 2.1) | 40 | -3 | (-3, -2) | 0.1 | 0.0 |
| Decision Memo for Positron Emission Tomography (FDG) for Infection and Inflammation (CAG-00382N) | 26 | 1.4 | (0.7, 2.3) | 26 | -3 | (-4, 0) | 0.0 | 0.0 |
| Decision Memo for Positron Emission Tomography (FDG) for Solid Tumors (CAG-00181R) | 34 | -0.4 | (-0.8, 0.2) | 34 | -3 | (-3.5, -2) | 0.0 | 0.0 |
| Decision Memo for Positron Emission Tomography (FDG) for Solid Tumors (CAG-00181R4) | 53 | -0.4 | (-0.7, -0.1) | 53 | -3 | (-4, -2) | 0.6 | 0.1 |
| Decision Memo for Positron Emission Tomography (N-13 Ammonia) for Myocardial Perfusion (CAG-00165N) | 39 | -0.1 | (-1.1, 0.6) | 39 | 2 | (0, 5) | 0.9 | 0.5 |
| Decision Memo for Positron Emission Tomography for Initial Treatment Strategy in Solid Tumors and Myeloma (CAG-00181R3) | 36 | -0.1 | (-0.8, 0.2) | 36 | -3 | (-3, -2) | 0.1 | 0.0 |
| Decision Memo for Screening for Colorectal Cancer - Stool DNA Testing (CAG-00440N) | 22 | 2.5 | (0.4, 11.1) | 22 | -1 | (-4, 3) | 1.2 | 0.4 |
| Decision Memo for Screening for Lung Cancer with Low Dose Computed Tomography (LDCT) (CAG-00439N) | 38 | 2.4 | (0.4, 6) | 38 | -2 | (-2.5, 0) | 1.1 | 0.5 |

| Decision Memo | | | | | | | | |
|---|---|---|---|---|---|---|---|---|
| Decision Memo for Screening for the Human Immunodeficiency Virus (HIV) Infection (CAG-00409N) | 32 | 1.5 | (0, 3) | 32 | -4 | (-4, -3) | 0.5 | 0.9 |
| Decision Memo for Serum Iron Studies (Addition of Restless Leg Syndrome as a Covered Indication) (CAG-00263R) | 60 | 1.2 | (0.5, 2.3) | 65 | 0 | (-2, 0) | 0.1 | 0.4 |
| Decision Memo for Sleep Testing for Obstructive Sleep Apnea (OSA) (CAG-00405N) | 59 | 1 | (0.6, 1.8) | 59 | 0 | (-2, 1) | 0.5 | 0.3 |
| Decision Memo for Stem Cell Transplantation (Multiple Myeloma, Myelofibrosis, and Sickle Cell Disease) (CAG-00444R) | 37 | 1.8 | (1.5, 2.9) | 38 | 1.5 | (-1, 3) | 0.8 | 1.2 |
| Decision Memo for Supervised Exercise Therapy (SET) for Symptomatic Peripheral Artery Disease (PAD) (CAG-00449N) | 48 | -0.4 | (-1.2, 0.3) | 48 | 0.5 | (-1, 2) | 0.4 | 0.8 |
| Decision Memo for Surgery for Diabetes (CAG-00397N) | 47 | 1.5 | (0.3, 3.3) | 47 | -1 | (-2, 0) | 0.2 | 0.1 |
| Decision Memo for Thermal Intradiscal Procedures (CAG-00387N) | 140 | -1.3 | (-1.5, -1) | 141 | -2 | (-2, -1) | 0.5 | 0.0 |
| Decision Memo for Transcatheter Aortic Valve Replacement (TAVR) (CAG-00430N) | 33 | 2.7 | (0.2, 6.2) | 33 | -4 | (-4, -3) | 0.1 | 0.1 |
| Decision Memo for Transcatheter Aortic Valve Replacement (TAVR) (CAG-00430R) | 141 | 1.6 | (1.2, 2.8) | 141 | -2 | (-2, -1) | 1.0 | 0.4 |
| Decision Memo for Transcatheter Mitral Valve Repair (TMVR) (CAG- 00438N) | 21 | 1.5 | (-0.5, 4.6) | 21 | -4 | (-4, -3) | 0.1 | 0.0 |
| Decision Memo for Transcutaneous Electrical Nerve Stimulation for Chronic Low Back Pain (CAG-00429N) | 48 | -0.3 | (-0.9, 0.1) | 49 | 1 | (-2, 4) | 0.1 | 0.0 |
| Decision Memo for Vagus Nerve Stimulation (VNS) for Treatment Resistant Depression (TRD) (CAG-00313R2) | 24 | 1.8 | (1, 4.5) | 24 | -1 | (-2.5, 2.5) | 0.9 | 0.3 |
| Decision Memo for Vagus Nerve Stimulation for Treatment of Resistant Depression (TRD) (CAG-00313R) | 37 | 1.8 | (-0.1, 2.4) | 38 | -4 | (-5, -2) | 0.8 | 0.1 |

| Memo | # Memo Articles w/ RCRs | Median ΔRCR | 95% CI | # Memo Articles w/ Years | Median ΔYear | 95% CI | $KLD_F$ | $KLD_{RO}$ |
|---|---|---|---|---|---|---|---|---|
| Decision Memo for Ventricular Assist Devices for Bridge-to- Transplant and Destination Therapy (CAG-00432R) | 43 | 4.7 | (0.4, 8.2) | 43 | -3 | (-4, -2) | 0.6 | 0.2 |
| Proposed Decision Memo for Erythropoiesis Stimulating Agents (ESAs) for Treatment of Anemia in Adults with CKD Including Patients on Dialysis and Patients not on Dialysis (CAG-00413N) | 577 | -1.1 | (-1.2, -0.9) | 597 | 5 | (4, 6) | 0.6 | 0.2 |

Notes. This table shows the median $\Delta RCR$ and $\Delta Year$ by memo (N = 68). For each memo, the number of memo articles with non-provisional RCR (# Memo Articles w/ RCRs) and publication year (# Memo Articles w/ Years) data are recorded. The same values are reported for the comparison set. $\Delta RCR$ and $\Delta Year$ values for each memo are reported along with the 95% confidence interval (bracketed numbers) as described in the Methods. Only those memos with at least twenty memo articles are shown. Kullback-Leibler Divergence ("KLD") columns are calculated as described in the Supplemental Methods with numbers closer to zero indicating a smaller divergence from a uniform distribution (e.g., all funders/research organizations support equal shares of articles cited in memo). KLD is reported for funders ($KLD_F$) and research organizations ($KLD_{RO}$).

**Supplemental Methods**

*Assigning award years for funding associated with memo articles*

Associating project (i.e., award) data with articles using PubMed and RePORTER is non-trivial. Often an abbreviation of the project number (e.g., the core project number) is reported, which can be associated with multiple project years. For instance, a scientist may receive a 3-year award and, at the conclusion of the first award, apply for a second 3-year award based on the results produced by the first award. If the scientist is successful, NIH appends a new suffix (e.g., "-02") to the core project number (e.g., "R01CA031770") assigned to the first project. Often, scientists will only report the core project number in a publication, which creates ambiguity when assigning a specific year to the cited funding. Similarly, RePORTER uses publication data reported in award applications to link articles to funding (NIH, 2024).

Unfortunately, inclusion of an article in an award application indicates the research was done prior to its being cited in that application. In both cases the year can be imputed by using award history data available in RePORTER.

I imputed the year for each award from an NIH Institute or Center (IC) that had records available in RePORTER (N = 2,444)[iv]. First, all projects (i.e., full project numbers) associated with a core project number were retrieved from RePORTER and linked to

---

[iv] This number is greater than the total number of unique projects funded by NIH ICs (N = 1,996) because the unique count only considers core project numbers (e.g., R01CA031770). For this analysis, full project numbers (e.g., R01CA031770-01) are imputed using the article publication date. Therefore, some core project numbers may be repeated (e.g., R01CA031770-01, R01CA031770-03, etc.) since they are cited by multiple publications from different years.

the memo article that cited their core project number. Each project was then compared to the citing article's publication year (article publication year - project year) and the project with the difference closest to one was assigned as the award year used in subsequent calculations described in the Methods section of the main paper. I used this procedure because I assumed that more recent awards were more relevant to an article that cited the associated core project number but that awards made in the same year as the publication or after it were less relevant (though not excluded). For awards that could not be linked to a record in RePORTER, I assumed that the award was issued the year prior to the publication of the article citing it.

*Measuring concentration in funders and research organizations by memo*

Data on the funders and organizations involved in the memo research ecosystem can also be used to measure whether few or many funders/research organizations support the articles cited by CMS memos. I use the Kullback-Leibler Divergence (KLD) to measure the divergence of the observed distribution of funders/research organizations from a uniform distribution (e.g., where all proportions are equal) using equation S1:

$$(S1) \quad KLD_x = \sum_i^N p_i \ln(N * p_i)$$

where $x$ is the individual memo, $i$ is the funder or research organization, $N$ is the total number of funders or research organizations supporting the memos cited in $x$, and $p_i$ is the proportion of articles supported by funder or research organization $i$ ($\sum_i^N p_i = 1$). $KLD_x$ values are approximations because proportions were calculated using the number articles with funder/research organization information in place of the total number of articles cited in a memo (many articles could not be linked to funder/research organization data). For articles citing multiple funding/research organizations, the count for the article is divided into equal fractions for each entity (e.g., each funder would receive 0.25 counts for a memo article supported by four different funders). The distribution of KLD for funders can be compared to that of research organizations using a paired Wilcoxon signed rank test.

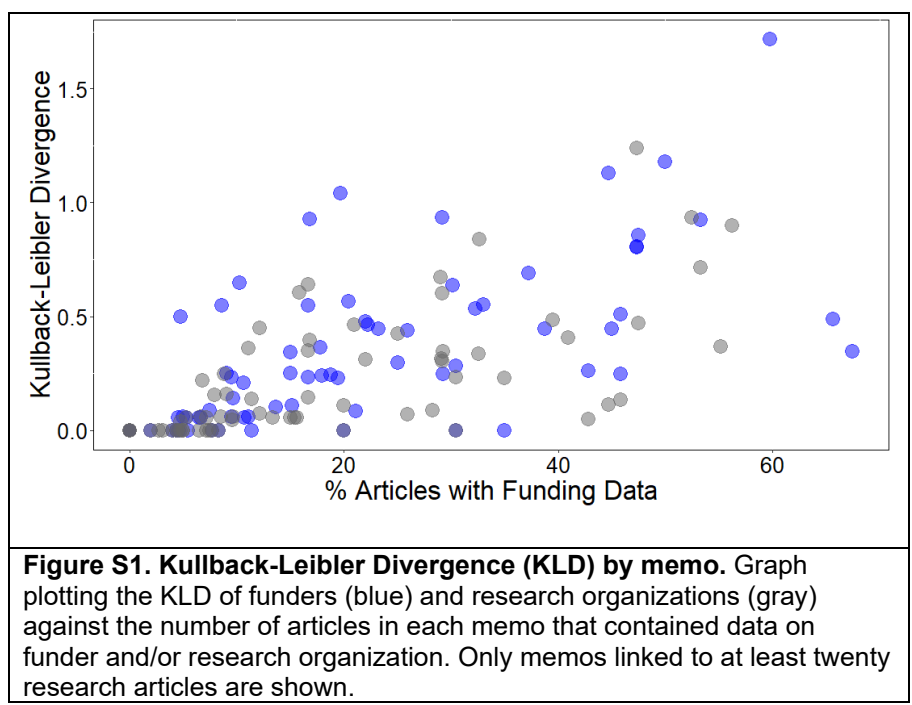

**Figure S1. Kullback-Leibler Divergence (KLD) by memo.** Graph plotting the KLD of funders (blue) and research organizations (gray) against the number of articles in each memo that contained data on funder and/or research organization. Only memos linked to at least twenty research articles are shown.

Figure S1 shows that memos take on a wide range of KLD values. The KLD distribution for funders is statistically different from that of research organizations (pseudo-median of the difference = 0.14; p-value = $2E^{-6}$), indicating that the distribution of support for memo articles among funders is more skewed than it is for research organizations. In other words, support for articles is distributed less equally among funders (i.e., some funders support a larger share of articles) than among research organizations.

*References*

NIH. (2024). *Research Portfolio Online Reporting Tools Expenditures and Results (RePORTER)* https://reporter.nih.gov/publications